\documentclass[a4paper]{jpconf}
\usepackage{graphicx}
\usepackage{xspace}
\newcommand{\Fermi}{\emph{Fermi}\xspace}

\begin{document}
\title{DIRAC framework evaluation for the \Fermi-LAT and CTA experiments}

\author{L Arrabito$^{1}$, J Cohen-Tanugi$^{1}$, R Graciani Diaz$^{2}$, F Longo$^{3}$, M Kuss$^{4}$, F Piron$^{1}$, M Renaud$^{1}$, V Rolland$^{1}$, M Sapunov$^{5}$, A Tsaregorodtsev$^{6}$ and S Zimmer$^{7}$}

\address{$^{1}$ Laboratoire Univers et Particules, Universit\'{e} de Montpellier II Place Eug\`{e}ne Bataillon - CC 72, CNRS/IN2P3, F-34095 Montpellier, France}
\address{$^{2}$ University of Barcelona, Diagonal 647, ES-08028 Barcelona, Spain}
\address{$^{3}$ Department of Physics, University of Trieste, via Valerio 2, Trieste and INFN, Sezione di Trieste, via Valerio 2, Trieste, Italy}
\address{$^{4}$ Istituto Nazionale di Fisica Nucleare, Sezione di Pisa, I-56127 Pisa, Italy}
\address{$^{5}$ formerly at Centre de Physique des Particules de Marseille, 163 Av de Luminy Case 902, CNRS/IN2P3, 13288 Marseille, France}
\address{$^{6}$ Centre de Physique des Particules de Marseille, 163 Av de Luminy Case 902, CNRS/IN2P3, 13288 Marseille, France}
\address{$^{7}$ The Oskar Klein Centre for Cosmoparticle Physics and Department of Physics, Stockholm University, AlbaNova, SE 106 91, Stockholm, Sweden}
\ead{arrabito@in2p3.fr}

\begin{abstract} 
DIRAC (Distributed Infrastructure with Remote Agent Control) is a general framework for the management of tasks over distributed heterogeneous
computing environments. It has been originally developed to support the production activities of the LHCb (Large Hadron Collider Beauty) experiment
and today is extensively used by several particle physics and biology communities. Current (\Fermi Large Area Telescope -- LAT) and planned (Cherenkov
Telescope Array -- CTA) new generation astrophysical/cosmological experiments, with very large processing and storage needs, are currently
investigating the usability of DIRAC in this context. Each of these use cases has some peculiarities: \Fermi-LAT will interface DIRAC to its own
workflow system to allow the access to the grid resources, while CTA is using DIRAC as workflow management system for Monte Carlo production and
analysis on the grid. We describe the prototype effort that we lead toward deploying a DIRAC solution for some aspects of \Fermi-LAT and CTA needs. 
\end{abstract}

\section{Introduction}
\label{intro}

The Large Area Telescope (LAT) is the primary instrument on the \emph{Fermi Gamma-ray Space Telescope} mission
, launched on June 11, 2008.
It is the product of an international collaboration between DOE, NASA and academic US institutions as well as international partners in France,
Italy, Japan and Sweden.
The LAT is a pair-conversion detector of high-energy gamma rays covering the energy range from 20 MeV to more than 300 GeV \cite{LATinstrument}.
It has been designed to detect gamma rays in a broad energy range, with good position resolution ($<$10 arcmin) and
an energy resolution of $\sim$10\%.
The LAT has been routinely monitoring the gamma-ray sky and has shed light on the extreme, non-thermal Universe.
A brief and recent review of \Fermi-LAT discoveries can be found in \cite{Thompson2013}.

The LAT response to gamma rays is parametrized by the so-called ``instrument response functions'' (IRFs), which together with the data from the
instrument are provided to the scientific community\footnote{Data release and software maintainance is done via the Fermi Science Support Center
  http://fermi.gsfc.nasa.gov.}.
As described in \cite{LATpass7}, IRFs are derived using Monte-Carlo (MC) simulations and also corrected for discrepancies observed between flight and
simulated data, as the LAT team gains insight into the in-flight performance of the instrument.
In the near future, major improvements are expected from the new ``Pass 8'' data, such as an increased effective area with respect to the current
``Pass 7'' public data \cite{LATpass8}.
These improvements correspond to a radical revision of the LAT event-level analysis.
The optimization of the event reconstruction and of the background rejection, and the full characterization of the new IRFs,
require the production of large simulated data sets including gamma rays and charged cosmic backgrounds (protons, heavy ions, electrons).
These simulations are also fundamental for high-level analyses which will require a proper evaluation of the residual backgrounds (e.g., the
extragalactic diffuse emission \cite{LATdiffuse} or the cosmic electron-positron spectra \cite{LATe+e-}).\\

The Cherenkov Telescope Array (CTA) project \cite{cta_overview} is the next generation of Imaging Atmospheric Cherenkov Telescopes (IACTs), operating
in the high- and very high-energy gamma-ray domain (between a few tens of GeV and 200 TeV). It will consist of two arrays of 50-100 telescopes of different sizes, located
in each hemisphere. The CTA consortium gathers more than 1000 scientists and engineers from more than a hundred institutions world-wide.
The project is currently in its preparatory phase. The construction is planned to be completed around 2018-2020.

During the current CTA preparatory phase, large computing and storage resources are needed mostly for MC studies. In particular, the selection
of the CTA sites (North and South) has a significant impact on the final sensitivity of the instrument. 
The CTA MC working group is studying the impact of these various parameters by means of detailed MC simulations of the detector response to extensive air showers. Large sets of simulated events are generated for different primary particles. 
Moreover, once CTA sites have been selected and the construction phase has started, more detailed simulations will be produced in order to test analysis algorithms and to determine the final performance of the instrument. \\

In order to fulfill present and future requirements for MC massive production and data analysis of \Fermi-LAT and CTA, we have proposed the use of the
EGI grid infrastructure and of the DIRAC (Distributed Infrastructure with
Remote Agent Control) \cite{DIRAC} framework for both current and future experiments. The DIRAC system, originally developed to support production activities of the LHCb experiment, today serves several communities. Compared to LHCb DIRAC installation, expanding over half a dozen powerful servers, the CTA installation is still rather modest, and it is currently being upgraded. The work presented in this paper was served by a DIRAC installation running on two virtual servers having, in total, 6 cores, 6 GB of RAM and 1.5 TB of local disk, plus a third machine hosting the web portal. \Fermi-LAT is using the French NGI multi-community DIRAC installation running on five servers \cite{FG-DIRAC}. In section \ref{Fermi} we describe the developments that have been necessary to extend the \Fermi-LAT pipeline to the grid through the DIRAC system. The context for CTA is quite different, since the project is in its preparatory phase and no existing production system was available for the management of the different computing activities. In section \ref{CTA} we present the work done to migrate both the CTA MC production and its analysis by the CTA physicists on the grid within the DIRAC framework. The first results from the MC campaigns in 2013, in terms of resource usage, are also presented. Section \ref{conclusions} is devoted to conclusions and perspectives for future work.

\section{Use of DIRAC for \Fermi-LAT massive simulations on the grid}
The data acquired by the \Fermi-LAT in orbit is transferred via satellite downlink several times per day, and then sent to an offline processing
system, which is hosted by the LAT ISOC (Instrument Science Operations Centers) based at the SLAC National Accelerator Laboratory in California. 
The \Fermi-LAT data processing pipeline (see, e.g., \cite{LATp2} for a detailed technical description) was designed with the focus on allowing the management of
arbitrarily complex workflows and handling multiple tasks simultaneously (e.g., prompt data processing, data reprocessings and science analysis).
We briefly recall its basic functionalities in section \ref{LAT_pipeline}. Then we describe its interface to DIRAC (section
\ref{LAT_DIRAC}) and we present the results of the first MC simulations which we performed on the grid during Summer 2013 (section \ref{LAT_results}).

\subsection{The data processing pipeline}
\label{LAT_pipeline}
Each pipeline task defines a workflow, specifying a sequence of procedures.
These procedures are then either executed directly on the workflow system or are delegated to job control daemons.
From the beginning, the system was tailored to be interfaced to local computing resources such as the batch farms at SLAC and the CNRS/IN2P3
computing center (CC-IN2P3 hereafter). To that extent, each job control service has to run as
a daemon on a host at the relevant site. Information is exchanged via remote method invocation between the job control service and the pipeline
server. Each task may contain an arbitrary number of streams or sub-streams, i.e. instances of a process. In the simplistic case of a MC production,
each stream creates a run containing a pre-defined number of simulated events. In case of a gamma-ray simulation, each stream simulates 5k events in
$\sim$90 minutes typically, and produces a small volume of data ($\sim$2.3 MB) when the full details on the event reconstruction are not saved.

\subsection{The Fermi-DIRAC setup}
\label{LAT_DIRAC}
\label{Fermi}
\begin{figure}[h]
\includegraphics[width=22pc]{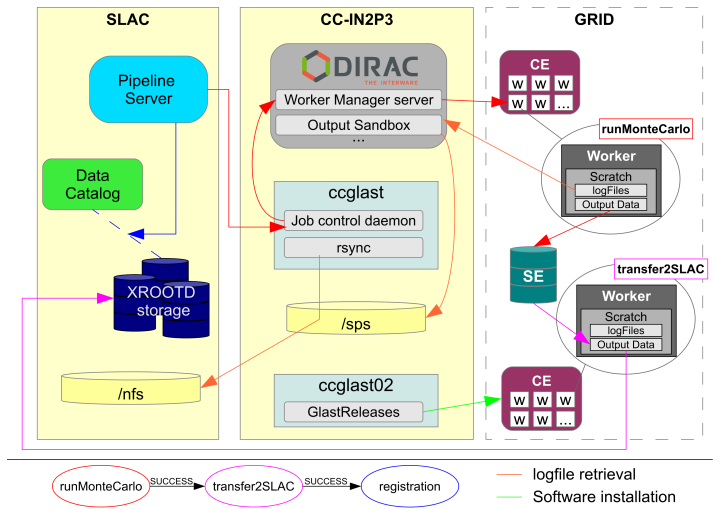}\hspace{2pc}
\begin{minipage}[b]{14pc}\caption{\label{LATegi}Schematic view of the \Fermi-LAT pipeline extension to the grid through the DIRAC system. Red arrows indicate the execution sequence of the MC simulation. Data and log file transfers are shown with magenta and orange arrows, respectively.
The blue arrow indicates the registration of data in the \Fermi data catalog. The LAT software installation on the grid sites is indicated by the green arrow.}
\end{minipage}
\end{figure}

The DIRAC interface to the data processing pipeline was designed to delegate computationally intense MC production to other sites.
It has been implemented using French NGI DIRAC servers at CC-IN2P3 (see Fig.~\ref{LATegi}), which run all DIRAC components, Workload Management System (WMS), Data Management System (DMS), agents and web portal, and host the MySQL databases. In addition, a \Fermi-specific DIRAC extension was developed through a dedicated package which provides basic API functionality in accordance with the one
used in the remaining pipeline system (see \cite{CHEP2012_SZ} for a detailed description). In particular, the extension provides wrappers to status
query and job submission.
The trigger to submit new jobs is done via the pipeline and its associated user interfaces. Thus, re-running failed jobs is done via the pipeline interface, which in turn results in a fresh re-submission of a DIRAC job with the exact same properties as the initial (failed) job. As a consequence, our current setup cannot fully utilize the possibilities to re-run failed jobs from within DIRAC.
At present, the DIRAC system employed by \Fermi-LAT does not support the means to request specific resources. As a consequence, jobs may be sent off to sites where the software is either not available or faulty. Instead of relying on discovery services such as those provided by the Berkeley Database Information Index (BDII), the \Fermi-DIRAC setup makes use of a custom software tag resource database. This is a separate service which relates a given site in DIRAC with the available software along with its current status. Statuses are queried automatically through an agent and jobs are matched to sites according to their requirements. Even if the matching becomes trivial for MC production, since the software releases are installed at all sites, this service is particularly useful during the setup of a production. Furthermore this system provides us with additional steering handles without the need to modify parts of the DIRAC configuration and or of the pipeline setup. 

The worker node that performs the actual computing step communicates with the pipeline through a series of emails, which in the case of local batch
systems provides a robust interface. In order to emulate this behaviour, the \Fermi-DIRAC setup utilizes the DIRAC notification layer that enables every
grid worker node to either send the required email directly or to communicate with the DIRAC server that then sends the relevant information on behalf of the
workers. Output data are stored on regular grid storage elements (StoRM, etc.) but picked up within the pipeline workflow system to be
transferred to XROOTD at SLAC (see Fig.~\ref{LATegi}). Once transferred, the data on grid storage elements are scheduled for removal. 

\subsection{First MC simulations on the grid}
\label{LAT_results}
A first working prototype of the grid/pipeline interface with DIRAC was completed in Summer 2013.
The first MC production on the grid consisted of 4000 streams, each one simulating 5k gamma-ray events.
This first test was limited to four EGI grid sites supporting the GLAST Virtual Organization (VO): CNAF, MSFG, OBSPM and BARI (sorted by
resource usage). They represent about one third of the GLAST VO resources.
The production used between 100 and 300 cores simultaneously, and it ran smoothly with a 97\% success rate.
The few failed jobs exhibited communication issues between the grid sites and the pipeline, which need to be
further investigated. 

\section{Use of DIRAC for CTA massive simulations on the grid}
\label{CTA}

The use of the EGI grid infrastructure for the MC production was introduced in CTA for the first time in 2008 with the creation of the CTA VO. The motivations which led to this choice were the following: large computing and storage requirements for the MC production
(see section \ref{MC_computing_needs}); the possibility of generating the required event statistics by splitting the production into independent
jobs and the software portability to the grid environment. Today, the CTA VO is supported by 21 EGI grid sites spread over 7 countries, with resources of
the order of several thousands of available logical cores and more than 650 TB of dedicated storage. During the first years, the grid infrastructure
has been exploited exclusively for the MC production activity. However, typical user analysis require the processing of about 30 TB, so that the
retrieval of a given dataset for further local processing is quite inefficient. In order to deploy the user analysis to the grid, thus reducing
latencies due to the data transfers, and in order to have an optimized usage of the grid resources, we started the migration of both the MC production
and the user analysis into the DIRAC system in 2011 \cite{CHEP2012_LA} (section \ref{CTA_DIRAC}). Finally the computing model applied to both
these activities is described in section \ref{CTA_CM}.

\subsection{MC computing needs}
\label{MC_computing_needs}

In order to estimate the computing and storage requirements for MC studies, we split the simulation chain into three main steps:
{\it 1)} shower generation and propagation in the atmosphere; {\it 2)} simulation of the telescope array response and {\it 3)} reconstruction and
analysis. These steps are executed sequentially and the output of each of them serves as input for the subsequent one. The first two steps, in which
the simulated data are produced, can be executed separately or within the same execution unit. This `production' activity is centralized and performed
by a team of managers, in charge of providing the CTA consortium with the products of step 2. In step 3, the `analysis' activity is performed by CTA
users and consists of several tasks.
The products of step 1, which represent 95\% of the overall data volume, are stored and conserved over periods of several months in view of
possible re-processing, i.e. the repetition of step 2 with different telescope configurations.
The products of step 2 should be conserved over longer periods of 1-2 years.

The necessary statistics for MC studies amounts to $\sim$${10}^{10}$ simulated events for different primary particles.
In order to collect such a statistics, each MC campaign is made of $\sim$200k jobs, each one producing 50k events.
The CPU time needed by a production is $\sim$8 M HS06 hours \cite{HS06}, 70\% of which is consumed by step 1.
Storage also represents a major challenge, since each MC campaign produces $\sim$550 TB of simulated data, including $\sim$30 TB to be analyzed by CTA
users in step 3.

\subsection{The CTA-DIRAC setup}
\label{CTA_DIRAC}
The DIRAC functionalities which motivated the migration of both MC computing activities (production and analysis) within this framework are the
following: optimization of the resource usage thanks to the implementation of the `pilot job' mechanism \cite{DIRAC}; easy customization for the
specific needs of a user community; central management of the different grid activities and of the VO policies; easy integration of heterogenous
resources; rich user interface (python API, command line and web portal). Moreover, DIRAC includes its own `native' catalog, the DIRAC File Catalog
(DFC), which covers all the functionalities of the LHC File Catalog (LFC), i.e. the `replica catalog' functionalities, offering at the same time the
support for user-defined meta-data. This last feature is particularly desirable in the context of the MC massive production. Indeed, each MC campaign
produces about 1 M files which must be organized in a logical structure and whose provenance must be tracked. Additionally, scientists should be able to easily select a given dataset according to some criteria relevant for their analysis. The DFC has a hierarchical structure where the meta-data key-value pairs are either affected to directories, or to files. Each sub-directory inherits the meta-data of its parent directory, while files inherit the meta-data of their containing directory. Finally, it is possible to declare some of the meta-data as `searchable', so that they are exposed for data searches.

The migration of the MC computing activities has been realized in two phases. The first phase consisted of the deployment of a DIRAC instance dedicated to CTA, which is composed of 4 servers hosted at PIC and CC-IN2P3 computing centers. The second phase consisted of the development of a CTA-specific DIRAC extension, called `CTA-DIRAC',
which includes the following modules: CTA software deployment on the grid; CTA workflow submission (production and analysis) and interface of the
production workflows to the DFC. All these modules make use of the DIRAC API to wrap the CTA applications and to manage the output data. In order to interface the production workflows with the DFC, we have first identified the meta-data that are relevant for searches (MC software version, particle flavour, CTA site altitude, telescope array configuration, etc.), as well as non-searchable ones (application exit status, DIRAC job identifier, etc.). Secondly, we have established a directory structure for the MC products, on the basis of the inheritance rules of the DFC. Finally, the DFC was interfaced with the production workflows, which extract the meta-data values, either from their input configuration parameters, or from run-time informations, to then fill the DFC.

\subsection{Computing model and resource usage}
\label{CTA_CM}

The CTA-DIRAC setup has been successfully exploited during the MC campaigns and analysis in 2013. The computing model currently used distinguishes two types of sites, i.e. Tier 1 (T1) and Tier 2 (T2), each having a particular role. The main difference between them is that while T2s provide only CPU, T1s are also providing support for long term storage of the data. With this difference in mind, T1s are employed for both MC production and analysis, while T2s are only used for MC production. Concerning the MC production, both T1s and T2s participate in terms of CPU, according to their capacity, but MC products are stored exclusively at T1s. For data access efficiency reasons, in our model jobs are input-data driven, so that only T1s are used for re-processing and for user analysis, thus acting as `analysis centres'. Taking into account the storage and CPU requirements for MC production and analysis, T1 must satisfy the following criteria: large storage capacity (at least 15-20\% of the total); CPU capacity proportional to the storage capacity and a good CPU-storage connectivity. Also, since T1s are directly exposed to users, they should satisfy high reliability requirements. T2 requirements are less stringent, and thus they are only required to provide a minimal CPU power and a sufficient connectivity to T1's storage elements to upload the simulation products. Given these requirements, the current CTA T1s are: CYFRONET, DESY and CC-IN2P3, while all other sites supporting CTA are considered as T2s. The overall storage capacity at T1s is of about 650 TB, which allows the storage of only one distributed replica of the MC products. The computing model just described is flexible, however; for example, in some cases T2s can be enabled to host a fraction of MC products, thus increasing the overall storage available.

\begin{figure}[h]
\begin{minipage}{18pc}
\centering
\includegraphics[width=14pc]{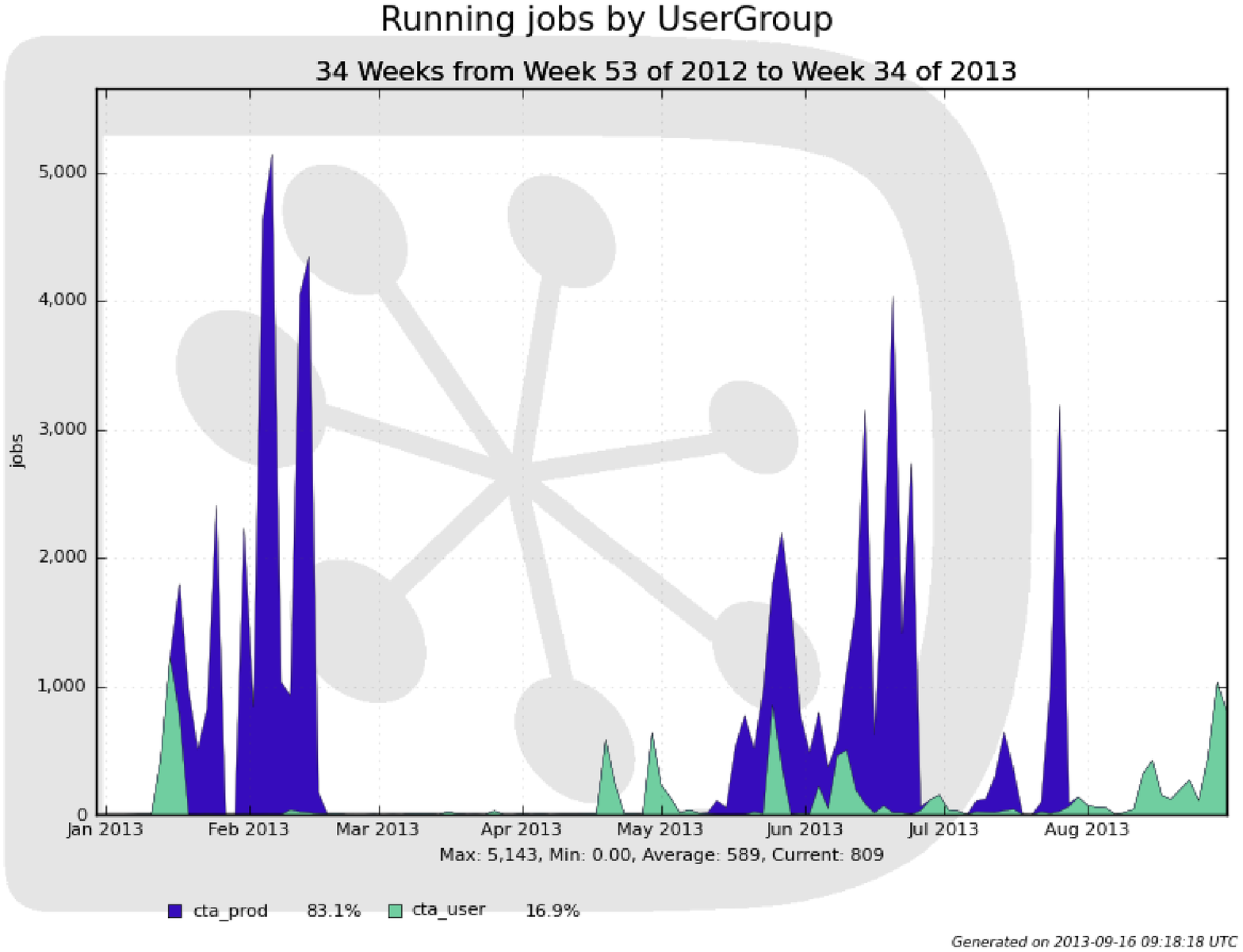}
\caption{\label{jobs}Number of concurrent running jobs in the period from January until September 2013. The blue and green colours correspond respectively to the MC production and the user analysis activities.}
\end{minipage}\hspace{2pc}%
\begin{minipage}{18pc}
\centering
\includegraphics[width=14pc]{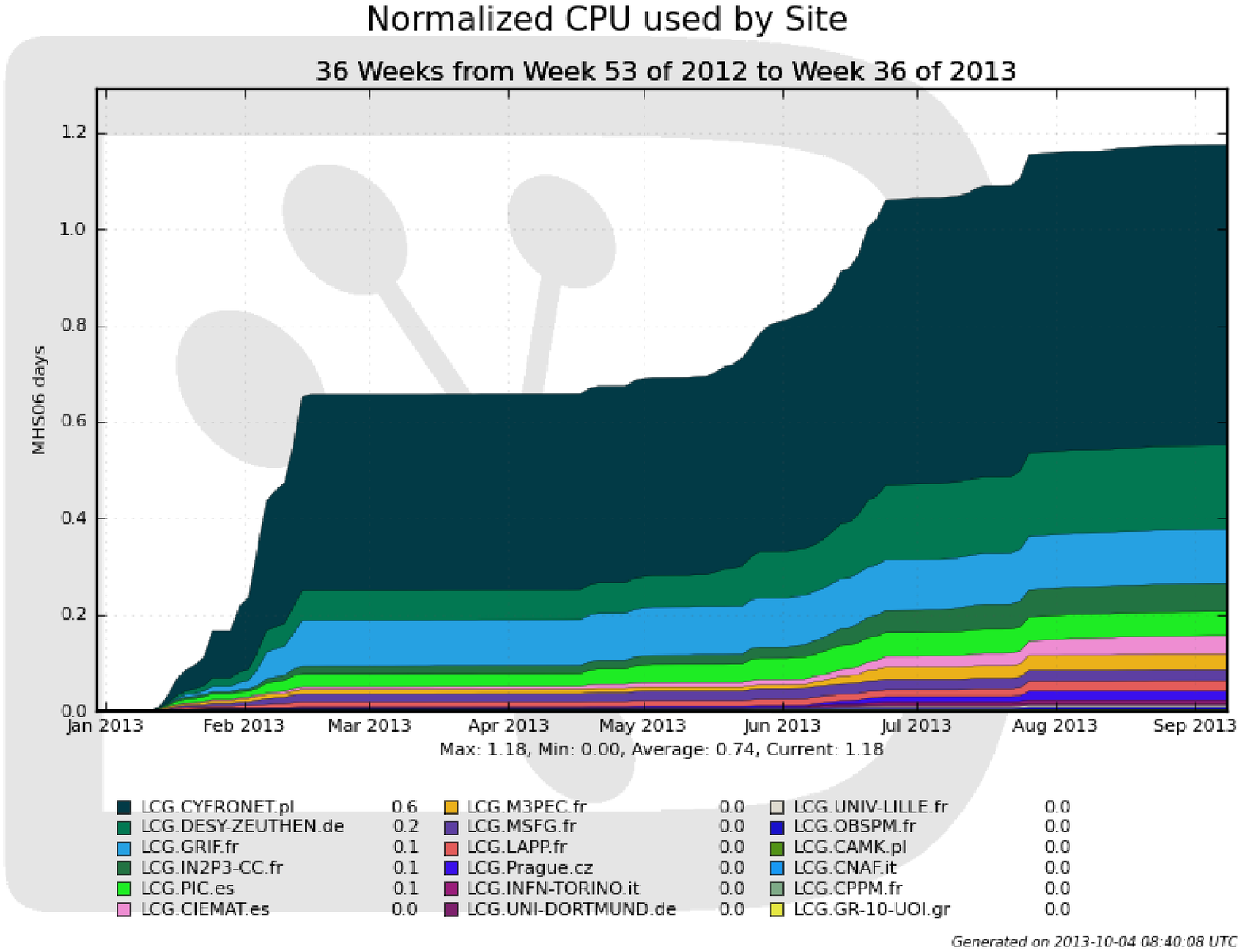}
\caption{\label{pie_CPU_used}Cumulative CPU consumption in the period from January until September 2013. The different colors refer to the grid sites where the jobs were executed.}
\end{minipage} 
\end{figure}

In the following, we report about the resource usage in the first 9 months of 2013. In Fig.~\ref{jobs}, we show the number of concurrent running
jobs in the considered period, where we distinguish the MC production in blue and the users' analysis in green. The peaks of activities correspond to
the different MC campaigns, during which we were able to achieve stable regimes of 4000-5000 jobs. This value is in good agreement with the resources expected to be
available, confirming the efficiency of the DIRAC WMS. The overall consumed CPU is about 30 M HS06 hours, with the site sharing shown in
Fig.~\ref{pie_CPU_used}. The main contributor is the CYFRONET T1, while the remaining 50\% of the consumed CPU is shared among ten other sites. The
overall volume of the data produced is about 600 TB, with a corresponding number of registered replicas in the DFC of about 3.3 M. The dataset selection,
based on DFC meta-data queries, can be done either through the DIRAC web portal, or through a CTA-DIRAC command line tool. The DFC response time
of the CTA-DIRAC instance is also adequate to the CTA usage, since typical user queries return a few tens of thousands of replicas and the average
response time is $\sim$8 sec per 10k replicas. Finally, it might be noted that about 13\% of the CPU time consumed is due to user analysis, with a
corresponding number of executed jobs of about 1 M.

\section{Conclusions and perspectives}
\label{conclusions}
In order to exploit the grid resources for the massive MC simulations and analysis of \Fermi-LAT and CTA, in both cases we have deployed a prototype setup, based on the DIRAC framework. The whole production chain of the \Fermi-DIRAC setup has been extensively tested, confirming that the DIRAC solution fulfils all the requirements imposed by the \Fermi-LAT pipeline. In the medium-term, we also plan to learn how the overall system behaves under stress through scalability tests, and to optimize the resource usage before entering production mode in view of the massive simulations of ``Pass 8'' backgrounds in Fall 2013. As for CTA, the DIRAC setup has been intensively exploited during the MC campaigns in 2013 and the subsequent performance studies for site candidate evaluation. The perfomance obtained shows that DIRAC is well adapted to both CTA production and analysis activities. Future developments will aim to further automate the management of the MC production, implementing automatic job and data operations according to predefined scenarios. 



\section*{References}
\begin{thebibliography}{99}
\bibitem{LATinstrument} Atwood W B et al. 2009 {\it The Astrophysical Journal} {\bf 697} 1071
\bibitem{Thompson2013} Thompson D J 2013 arXiv:1308.1870 
\bibitem{LATpass7} Ackermann M et al. 2012 {\it The Astrophysical Journal Supplement Serie} {\bf 203} 4
\bibitem{LATpass8} Atwood W B et al. 2013 {\it 2012 Fermi Symposium: eConf Proceedings C121028} arXiv:1303.3514
\bibitem{LATdiffuse} Abdo A A et al. 2010 {\it Physical Review Letters} {\bf 104} 101101
\bibitem{LATe+e-} Ackermann M et al. 2012 {\it Physical Review Letters} {\bf 108} 011103
\bibitem{cta_overview} Actis M et al. (CTA Consortium) 2011  {\it Experimental Astronomy} {\bf 32} 193
\bibitem{DIRAC} Tsaregorodtsev A et al. 2008 {\it Journal of Physics: Conference Series} {\bf 119} 062048
\bibitem{FG-DIRAC} Arrabito L et al. http://hal.archives-ouvertes.fr:hal-00766084 
\bibitem{LATp2} Dubois R 2009 {\it ASP Conference Series} {\bf 411} 189
\bibitem{CHEP2012_SZ} Zimmer S et al. 2012 {\it Journal of Physics: Conference Series} {\bf 396} 032121
\bibitem{CHEP2012_LA} Arrabito L et al. 2012 {\it Journal of Physics: Conference Series} {\bf 396} 032007
\bibitem{HS06} http://hepix.caspur.it/benchmarks/doku.php (last access date: $27^{th}$ Oct. 2013)

\end{thebibliography}
\end{document}